# Photon Management in Silicon Photovoltaic Cells: A Critical Review

Mohammad Jobayer Hossain[1,5,6,*], Mengdi Sun[1,7], Kristopher O. Davis[1,2,3,4]

[1]*CREOL, the College of Optics and Photonics, University of Central Florida, Orlando, Florida, USA*
[2]*Resilient Intelligent Sustainable Energy Systems (RISES) Faculty Cluster, University of Central Florida, Orlando, Florida, USA*
[3]*Department of Materials Science and Engineering, University of Central Florida, Orlando, Florida, USA*
[4]*FSEC Energy Research Center, Cocoa, Florida, USA*
[5]*American Institute for Manufacturing Integrated Photonics (AIM Photonics), Albany, New York, USA*
[6]*The Research Foundation for State University of New York, Albany, New York, USA*
[7]*Bradley Department of Electrical and Computer Engineering, Virginia Tech, Arlington, Virginia, USA*

[*]*Corresponding author: jobayer@knights.ucf.edu*

**Abstract**

With the practical efficiency of the silicon photovoltaic (PV) cell approaching its theoretical limit, pushing conversion efficiencies even higher now relies on reducing every type of power loss that can occur within the device. Limiting optical losses is therefore critical and requires effective management of incident photons in terms of how they interact with the device. Ultimately, photon management within a PV cell means engineering the device and constituent materials to maximize photon absorption within the active semiconductor and therefore reduce the number of photons lost through other means, most notably reflection and parasitic absorption. There have been great advancements in the front and the rear side photon management techniques in recent years. This review aims to discuss these advancements and compare the various approaches, not only in terms of increases in photogenerated current, but also their compatibility with different PV cell architectures and potential trade-offs, like increased surface recombination or scalability for high-volume manufacturing. In this review, a comprehensive discussion of a wide variety of the front and the rear side photon management structures are presented with suggestions to improve the already achieved performance further. This review is unique because it not only presents the recent development in photon management techniques, but also offers through analysis of these techniques and pathways to improve further.

*Keywords:* Light trapping, path length enhancement, photon management, recombination, silicon, PV cell

**Highlights**:

- A thorough exploration of diverse front and rear photon management structures is presented.
- Suggestions for further improving the already achieved cell performance are offered.
- Explains underlying physics and material properties dictating various energy conversion losses in PV cells.
- Explores optical benefits of photon management structures and their impact on recombination and resistive losses.
- This unique review discusses recent advances in photon management and provides in-depth analysis with pathways for further improvement.

**Nomenclature**:

1. PV: photovoltaic

2. LCOE: levelized cost of energy
3. PERC: passivated emitter rear contact
4. SHJ: silicon heterojunction
5. HIT: heterojunction with intrinsic thin-layer
6. POLO: polycrystalline silicon on oxide contacts
7. TOPCon: tunnel oxide passivated contact
8. IBC: interdigitated back contact
9. ARC: antireflection coating
10. SN: serial number
11. EQE: external quantum efficiency
12. DRIE: deep reactive ion etching
13. WGM: hispering-gallery modes
14. Al-BSF: aluminum back surface field
15. DWCNT: double wall carbon nanotube
16. HJ-IBC: heterojunction interdigitated back contact
17. TCO: transparent conducting oxides

## 1. Introduction

Photovoltaic (PV) energy conversion has now become one of the cheapest sources of electricity [1], less expensive than most fossil fuel-based resources. Sunlight is abundant on earth, and PV cells and modules directly convert incident photons into electricity using a process called photovoltaic effect. A wide variety of materials can be used to make PV cells, including organic semiconductors, perovskites, III-V semiconductors, chalcogenides, and of course silicon (Si). Even though some of these materials are less expensive to produce and others yield higher conversion efficiencies, crystalline Si (c-Si) has been and continues to dominate the PV manufacturing sector with a current market share of approximately 95% [2]. Si has a bandgap of 1.12 eV, which is an optimal value for a single junction PV cell based on the AM 1.5G solar spectrum. However, it has an indirect bandgap that results in weak absorption near the band edge. Other materials have direct bandgaps of a similar magnitude, so why does silicon remain the dominant material? For starters, Si is the second most abundant element on earth, behind only oxygen. Secondly, many decades worth of research, development, and production in the integrated circuits sector have led to an incredible understanding of how to work with Si and produce it at a large scale. Finally, Si-based modules have shown to be very durable and reliable over time with very low degradation rates reported from the field [3–7]. This, combined with the relatively high efficiencies obtained and the low production costs, have all contributed to the continued use of Si by PV manufacturers.

Incident photons coming from the sun serve as the "fuel source" in a PV cell. The wavelengths of these photons span the ultraviolet, visible, and infrared domains. Not all the incident photons enter the cell. As illustrated in Figure 1, some are reflected off the front surface of the cell and others are parasitically absorbed by other front surface layers of the device. Of the photons that do enter the semiconductor, not all will get absorbed and generate an electron-hole pair.

Photons with energies smaller than the bandgap either don't get absorbed at all or get absorbed and converted into heat (e.g., free carrier absorption). Others may reach the rear side of the cell and get transmitted through the device or absorbed by other rear surface layers (e.g., metal contacts). Even photons that are internally reflected at the rear surface can escape out the front. Some of these losses associated with the rear side of the cell could be addressed by making the Si substrate much thicker, but cost constraints prevent, and to a lesser extent bulk carrier recombination, prevent the use of very thick wafers.

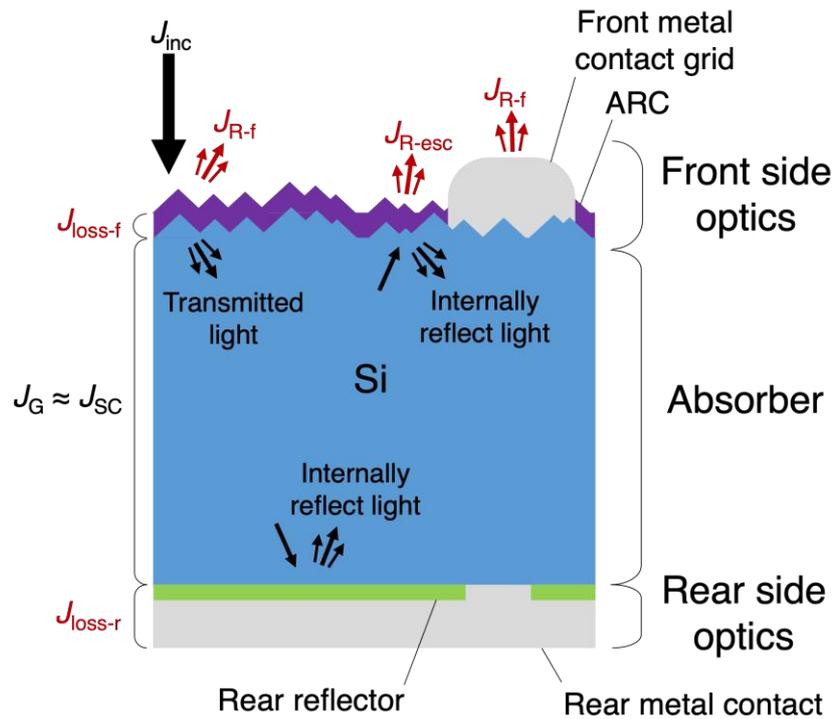

Figure 1: Simplified illustration of the light propagation in a silicon photovoltaic cell. The figure highlights the losses in red color, including reflection and absorption losses. The legend provides a breakdown of these losses, including loss in current density due to front reflectance ($J_{R-f}$), loss in current density due to escape reflectance ($J_{R-esc}$), loss in current density (absorption) at the front side ($J_{loss-f}$), and loss in current density (absorption) at the rear side ($J_{loss-r}$). This visualization provides valuable insights into the mechanisms that affect the efficiency of the photovoltaic cell, helping to identify areas for optimization and improvement.

Silicon, being an indirect band-gap material, experiences a rapid decrease in its absorption coefficient as the wavelength of the incident light approaches the band gap energy. (i.e., 1.12 eV or 1107 nm). This means that the thickness required to absorb all the photons increases rapidly with wavelength. The photon management structures play a crucial role in this scenario. In the presence of these structures, photons that are not initially absorbed by silicon in a single pass internally reflect back and forth between the front and rear sides of the cell (Figure 1), which enhances the effective optical path. Termed 'path length enhancement', this mechanism is characterized by the ratio between the effective optical path and the physical thickness, known as the path length enhancement factor.

The efficiency of a PV cell is now one of the driving factors that governs the levelized cost of energy (LCOE). Cell manufacturing costs have gotten so low that a larger fraction of the total PV system costs can now be attributed to things like the glass and aluminum (Al) needed to make the modules, the mounting hardware, the wiring and conduit, and the installation labor. Many of these factors scale with area, so an increase in cell efficiency can yield a lower cost per watt at the system level [8]. The current world record for silicon PV cell efficiency is 26.8% [9, 10] using a heterojunction structure, while the theoretical limit of such a cell, known as the Shockley-Queisser limit, is approximately 30% [11] under AM 1.5G solar spectrum. By implementing photon management techniques and minimizing losses such as recombination, resistive and reflection losses, the gap between the current efficiency levels and the Shockley-Queisser limit can be significantly reduced.

Recent advancements in photon management techniques for Si PV cells have led to improved theoretical understanding, fabrication methods, and cost reduction. The techniques range from nano-texturizing the active material to using resonators of various materials for front-side light trapping, and diffraction grating, planar metallic and dielectric layers for the rear side. These methods effectively trap light, resulting in an increase in photogenerated current density. However, these structures often come with drawbacks such as high surface recombination loss, which leads to low open-circuit voltage and fill factor, complex fabrication processes, higher contact resistivity and additional implementation costs, making them less suitable for industrial PV cell manufacturing. While many of these novel designs show promise in terms of photogenerated current density only, other performance parameters have yet to be fully explored.

Recent advancements in light trapping structures have led to a growing need for a comprehensive review of photon management in silicon PV cells within the research community. In our search for such papers, we have found several review papers on the topic, including those focusing on nanoscale photon management in silicon PV [12–14], nanostructured silicon PV [15], and thin silicon PV cells [16]. While these papers provide thorough analysis of different structures, they lack an examination of the various loss mechanisms and implementation feasibility associated with these structures. Additionally, many of these articles only focus on photon management from one side of the cell, either the front or rear. In this article, we aim to provide a comprehensive review of existing photon management schemes, with a particular emphasis on both the front and rear sides of the cell.

Silicon PV cells are diverse both in terms of how they are designed and manufactured [17–20]. This variety takes the form of different cell architectures, etching and surface preparation processes (e.g., anisotropic wet texturing [21–23], isotropic wet texturing [24–27], surface cleaning [28–34]), doping processes [35–40], materials and deposition methods for passivation layers and optical coatings [41–46], and various metallization and manufacturing processes [47–55]. The major cell architectures include: aluminum back surface field (Al-BSF) cells [56, 57], the dominant architecture until recently; passivated emitter rear contact (PERC) cells [58–62], the newly emerged dominant architecture in the global market [63, 64]; silicon heterojunctions (SHJ) or heterojunction with intrinsic thin-layer (HIT) cells [65–70], a high efficiency technology that has a smaller market share due to higher manufacturing costs; and polysilicon-based passivating contact cells, a concept introduced many years ago [71], but recently attracting a lot of attention and development under the name of polycrystalline silicon on oxide contacts (POLO) [72–74], monofacial poly-Si contacts (monoPoly) [75, 76], or most frequently as tunnel oxide passivated contacts (TOPCon) [77–86]. The HIT and poly-Si architectures both feature carrier-selective, passivating contacts aimed at suppressing contact recombination. Other carrier-selective, passivating contact technologies exist and have their own advantages and disadvantages, but are not the focus of this review as they are still in the early stages of development. Additionally, interdigitated back contact (IBC) cells [87–89] are another cell structure that eliminates the use of metallization on the front side of the cell, thereby reducing the optical loss associated with light being reflected off these metal contacts. The IBC concept can be combined with other approaches, including heterojunctions and polysilicon-based passivating contacts to form very high efficiency cells, as in the current record holding 26.8% cell that is an IBC heterojunction [9, 10, 90].

In this article we consider the major photon management schemes, from the more conventional approaches, like the use of antireflection coatings (ARC), to more novel and recently developed approaches for these various cell architectures. Section 2 explains the underlying physics and materials properties that dictate the various energy conversion losses in a cell. These are categorized as optical (i.e., photons lost), recombination (i.e., charge carriers lost through recombination), resistive (e.g., voltage drop via inefficient charge transport), and selectivity, an issue typically limited to heterojunctions. Section 3 provides an up-to-date review of the major photon management approaches in terms of optical benefits, as well as their impact on recombination and resistive losses when incorporated into cells. Section 4 presents a comprehensive and objective analysis of the optical performance of the various schemes using optical simulations of the reflectance and quantum efficiency curves. Finally, Section 5 presents the overall conclusions.

## 2. Device Performance Parameters and Loss Mechanisms in Photovoltaic Cells

### 2.1. Device Performance Parameters

In addition to the cell architecture selected, the efficiency of a c-Si PV cell depends on many factors, including: (1) design choices, in terms of the physical layout and dimensions of constituent components of the device (e.g., front and rear dielectric layers, metallization); (2) the materials selected to form those constituent components; and (3) the materials processing steps used to fabricate the device. Materials selection and processing dictate the composition, structure, and physical properties of the constituent components that make up the PV cell. The physical properties ultimately govern the efficiency ($\eta$), which is measured by collecting the current or current density versus voltage ($I-V$ and $J-V$, respectively) of the device under standard test conditions (STC[1]). The illuminated $J-V$ curve of c-Si PV cell can be represented fairly well as a one diode equivalent circuit using the following simplified Shockley diode equation [92] that assumes a unity ideality factor ($n = 1$) and infinite shunt resistance ($R_{SH} = \infty$):

$$J = J_G - J_0 \; exp\left(\frac{V + JR_S}{kT} - 1\right). \tag{1}$$

Here, $J$ is the current density flowing from the PV cell to the external circuit, $V$ is the voltage at the terminals of the PV cell, $J_G$ is the photogenerated current density, $J_0$ is the dark saturation current density, $R_S$ is the series resistance of the PV cell, $k$ is the Boltzmann constant, and $T$ is the PV cell temperature[2]. The point of the $J-V$ curve where the product of $J$ and $V$ are maximum defines the maximum power current density ($J_{MP}$) and voltage ($V_{MP}$), and it is this point that defines the overall efficiency since it represents the power output of the device and the irradiance the input power. Some of the other key features of the illuminated $J-V$ curve, often called PV cell performance parameters, include the open-circuit voltage ($V_{OC}$, i.e., $V$ at $J = 0$), the short-circuit current density ($J_{SC}$, i.e., $J$ at $V = 0$), and the fill factor ($FF$), which is simply $\frac{J_{MP}V_{MP}}{J_{SC}V_{OC}}$.

The efficiency can be expressed directly by $V_{OC}$, $J_{SC}$, and $FF$ via the following equation, so maximizing all three is critical:

$$\eta = \frac{V_{OC} \; J_{SC} \; FF}{G_{STC}}, \tag{2}$$

where $G_{STC}$ is the irradiance at STC and equal to $100 \frac{mW}{cm^2}$. Beyond the maximum power point and efficiency, expressing performance in terms of these additional parameters can yield more insight into why a PV cell is performing well or not.

---

[1] Irradiance, $G_{STC}$ = 1,000 $W/m^2$ = 100 $mW/cm^2$; Spectral distribution defined as air mass 1.5 global (AM1.5G) [91]; Cell operating temperature, $T_{STC}$ = 25° C

[2] The product of $k$ and $T$ is the thermal voltage and at STC, where T = 25° C, $kT \approx 26$ meV.

For example, $J_G$ and $J_{SC}$ are typically equivalent for well-performing c-Si PV cells, and these parameters are strongly dependent on optical losses, as well as some recombination losses, in the PV cell. Because of this, $J_G$ and $J_{SC}$ are often the focus of many research articles reporting new photon management approaches. $J_G$ and $J_{SC}$ can be expressed by the following equation:

$$J_G = J_{SC} = e \int [1 - R(\lambda)] \Phi_{inc}(\lambda) IQE(\lambda) d\lambda \quad (3)$$

Here $\Phi_{inc}$ is the incident photon flux, IQE is the internal quantum efficiency, and $R$ is the fraction of reflected light. $R$ is the summation of both front reflectance ($R_f$) and the escape reflectance ($R_{esc}$), and both refer to the direct loss of photons (i.e., purely optical loss). $R_{esc}$ refers to light that enters into the cell through front side, is not absorbed by Si, becomes reflected from the rear side, is not absorbed by Si again as it propagates back up through the Si, and then exits the cell through the front side. For a typical silicon PV cell, the total $R$ can be measured as a function of $\lambda$ using a spectrometer, and then $R_f$ and $R_{esc}$ can be decoupled by looking at the increase in $R$ near the band edge, where light is weakly absorbed because of the indirect bandgap of Si [93, 94]. These lost photons can be expressed in terms of lost current by multiplying each parameter by $\Phi_{inc}$ and integrating over the relevant wavelength range in a manner similar to the equation for $J_G$ [57, 95]. The resulting current densities, $J_{R-f}$ and $J_{R-esc}$ shown in Figure 1, provide a quantitative measure of the impact of each on the resulting cell performance [57, 96–99]. The IQE term includes both optical and recombination losses. Parasitic optical absorption does not show up in $R$, but nonetheless is due to a loss of photons that ultimately reduces $J_G$ and $J_{SC}$. Recombination also affects the collection efficiency of carriers that are actually generated via optical absorption, but recombine before that can be extracted by the contacts.

The $V_{OC}$ is particularly sensitive to recombination losses, and it is here where cell architectures featuring passivation, carrier-selective contacts and high quality, monocrystalline substrates really set themselves apart. Because Si is indirect bandgap, the radiative carrier lifetime is quite long (e.g., in excess of hundreds of milliseconds), so Auger and Shockley-Read-Hall (SRH) recombination losses tend to be the mechanisms that must be considered carefully when designing and fabricating c-Si PV cells. Auger recombination is strongly dependent on carrier concentration, so limiting the doping concentration as much as possible tends to be common the course of action, although this must be balanced with resistive losses. SRH recombination is ultimately the bigger concern for most PV cells. This refers to nonradiative recombination through deep-level defect states present at the surfaces and in the bulk.

$V_{OC}$ can be expressed in various ways. By setting $J = 0$ in Equation 1, the relationship between $V_{OC}$ and $J_0$ becomes evident:

$$V_{OC} = \frac{kT}{e} \left( \frac{J_{SC}}{J_0} + 1 \right). \quad (4)$$

$J_0$ is therefore an important figure of merit for PV cells. Minimizing SRH recombination at the surfaces and in the bulk lowers $J_0$, thereby increasing $V_{OC}$. The so called implied $V_{OC}$ ($iV_{OC}$) is another way to express the $V_{OC}$ in terms of the excess carrier concentration ($\Delta n$):

$$iV_{OC} = \frac{kT}{e} \frac{(N_{A/D} + \Delta n)\Delta n}{n_i^2}. \quad (5)$$

where $N_{A/D}$ refers to the background acceptor or donor doping concentration and $n_i$ the intrinsic carrier concentration of the of the c-Si absorber. The relationship between $iV_{OC}$ and $\Delta n$ can be very useful, as it can provide a means of related carrier-related measurements like photoconductance and photoluminescence to the PV cell's $V_{OC}$ and conversely illumination-dependent $V_{OC}$ measurements (i.e., Suns-$V_{OC}$) to the injection-dependent excess carrier concentration and carrier recombination lifetimes [100, 101]. Furthermore, it shows that the recombination losses effectively set a ceiling on the final $V_{OC}$ that one can obtain. The $V_{OC}$ cannot exceed $iV_{OC}$, and in most cases $V_{OC}$ equals $iV_{OC}$. However, poor carrier selectivity of the electron and/or hole contacts can lead to situations where additional voltage is lost (i.e., $\Delta V = iV_{OC} - V_{OC} > 0$) [102–107]. This is particularly an issue with poorly performing heterocontacts without a homojunction present.

Finally, $FF$ is primarily associated with the resistive elements of the equivalent circuit, particularly $R_S$. $R_S$ is essentially a lumped parameter that includes many different resistive components, including the (ideally) asymmetric carrier transport to the electron and hole contact regions, charge transfer between the doped regions or TCOs to the metal contact grid or layer, current flow through the metallization, and eventually through the rest of the circuit. $FF$ is also dependent on recombination losses [108] and carrier selectivity losses [107].

## 2.2. Design Considerations and Losses at the Front Side

If we consider how the cell architecture, design, materials, and resulting properties influence the overall performance, it is helpful to break the cell into the front side, bulk absorber, and rear side, as in Figure 1. At the front, critical design elements include the surface morphology, the thickness and physical properties of any thin film layers, and dimensions, shape, and electrical properties of the front metallization grid. The surface morphology, including the size and shape of the surface texture, can impact the front surface reflectance ($J_{R-f}$), as well as the direction light will propagate within the c-Si absorber. This is covered in more detail in subsequent sections.

Thin films at the front surface can minimize $J_{R-f}$ when used as a ARC or multi-layer ARCs [109–114], when a wide bandgap material is used with the appropriate refractive index and thickness. It can also increase $J_{loss-f}$ if the materials has a significant absorption in the wavelengths of interest. This is particularly an issue for heterojunction cells (e.g., HIT) featuring doped amorphous silicon films and a TCO, like indium tin oxide (ITO) [115]. Parasitic optical absorption in the amorphous silicon layers primarily affects blue part of the spectrum, but the TCO can affect longer wavelengths and is highly dependent on the carrier concentration. The use of alternative passivating, carrierselective contact materials with wider bandgaps and therefore less absorption has been an area of active research with various metal oxides of particular interest [102, 103, 116–136]. For both homojunction and heterojunction devices, the thin film in direct contact with the front of c-Si absorber should lower surface recombination by passivating interface defects that act as SRH recombination centers and preferably also lowering the minority carrier concentration at the surface [46]. For cell architectures featuring a front homojunction (e.g., Al-BSF, PERC, TOPCon), this film is typically an insulating dielectric film, whereas for heterojunctions, the film must either be conductive or thin enough to still allow the desired charge carrier to be transported through the layer.

Finally, the front metallization grid impacts the optical, recombination, and resistive losses. Optical shading from the front metallization leads to front reflectance loss ($J_{R-f}$) and is dependent on the fractional area covered with metal and the shape of the metal grid lines. Similarly, metal-silicon interfaces are have a very high interface defect density, and tend to yield excessively high $J_0$ values [137–142]. Limiting the metal contacted area seems to be prudent, but limiting it too much can lead to poor current collection and significant voltage drop at the interface [141, 143, 144] and along the grid lines, thereby increasing $R_S$ and lowering $FF$. By printing narrow grid lines with a high aspect ratio, this can be mitigated to some extent. IBC cells obviously avoid some of these issues and trade-offs all together, but at come at a higher cost and added manufacturing complexity.

*2.3. Design Considerations and Losses in the Bulk Absorber*

Compared to the front and rear side, the design considerations of the bulk absorber are much simpler. Arguably the most desirable characteristic of the bulk absorber is to maximize the bulk carrier lifetime, because this sets a ceiling on the overall effective carrier lifetime of the entire device which is ultimately tied to the key recombination-related device parameters like $J_0$ and $V_{OC}$. The bulk carrier lifetime is strongly depends on the concentration of deep-level SRH defects formed by the presence of crystallographic defects within the Si lattice. This can be various point defects (e.g., impurities) and structural defects like grain boundaries and dislocation clusters. In years past, this was particularly an issue for cells made from multicrystalline (i.e., polycrystalline) silicon wafers. However, in recent years, the industry has almost completely transitioned to the use of monocrystalline silicon wafers grown using the Czochralski (Cz) method. Even with the use of Cz wafer, the dopant type (e.g., *n*, *p*), dopant species (e.g., B, Ga, P), dopant concentration, and absorber thickness remain important design considerations. The as-grown quality of the resulting crystals and subsequent processing steps, particularly those occuring at higher temperatures and/or those that introduce significant hydrogen to the absorber are also very critical. For this review, the thickness is of particular importance because the ability to absorb the majority of the band edge photons depends on the optical path length, which is based on both the thickness and the light trapping performance of the device. As the absorber thickness gets smaller due to cost pressure associated with the crystal growth process, light trapping becomes more and more critical.

*2.4. Design Considerations and Losses at the Rear Side*

In terms of recombination and resistive losses, the rear side shares similar design considerations as the front. To help keep $J_0$ low and $V_{OC}$ high, the rear surface should be well-passivated with limited contact between the rear metallization and the c-Si absorber. Similarly, this must be balanced with the resistive losses associated with limiting the rear contact fraction. Of course, the use of a passivating, carrier-selective contact stack at the rear can alleviate some of this tradeoff, as is the case with HIT and TOPCon cells. With regards to optical losses, the situation is quite different. Rather than limiting the front surface reflectance and broadband parasitic absorption, the primary optical considerations at the rear are to maximize the internal back reflectance [93, 94, 145, 146] and, if possible, redirect the internally reflected light to increase the optical path length and potentially help maintain a higher reflectance for subsequent internal reflection (e.g., at the front).

Both the rear surface morphology and the properties and thickness of the rear reflectors govern the internal reflectance and the light redirection. The rear reflectors should have a low refractive index for wavelengths near the band edge, since Si has a large refractive index ($n \approx 3.5$) and a large difference in the refractive index is what yields a large internal reflectance. For random upright pyramids formed via anisotropic etching at the front side and common low index rear reflectors like silicon oxide and silicon nitride at the rear (Figure 1), much of the light entering the c-Si absorber is transmitted at an angle in excess of the critical angle for total internal reflection. However, the rear reflector film thickness must be kept sufficiently thick (e.g., greater than 100-150 nm) to prevent energy transfer from the "transmitted" evanescent wave to the underlying metal contact [94]. The use of thinner films and lossy metals can lead to significant parastic absorption at the rear side leading to current loss in the form of the $J_{loss-r}$ term shown in Figure 1.

An important consideration for both the front and rear surface is that the surface morphology of the c-Si absorber can also affect the surface recombination. Planar surfaces are much easier to passivate, whereas nanostructured surfaces can be more of a challenge. Research focused on photon management can often focus on the optical performance of novel nanostructured surfaces without consideration of how effectively the surface can be passivated.

That isn't to say it is impossible, in fact aluminum oxide deposited by atomic layer deposition has been shown to provide some level of passivation of nanostructured black silicon [147–156].

## 3. Photon Management Structures used in Si PV

Incomplete light trapping and parasitic optical absorption are two major challenges in reaching the Shockley–Queisser efficiency limit. Various photon management techniques have been studied to address these issues, including reducing losses, enhancing the optical path, and increasing the probability of photon absorption. Figure 2 illustrates the performance parameters and the photon management structures of numerous silicon PV cells reported recently, carefully chosen to represent most of the cell types, photon management structures and surface passivation conditions. Here, the overall efficiency ($\eta$) is represented by the colorbar. The $V_{OC}$ and $J_{SC}$ of the cells are shown in X and Y axis respectively, and their performance parameters are listed in Table 1. The shape of the data points in the figure represents the front side photon management structures, and a white circle inside some of the data point means that their rear side is structured (i.e., not planar). The label of the data points includes the cell type and their serial number (SN) on Table 1. The vertical and horizontal dotted lines represent the maximum $V_{OC}$ and $J_{SC}$ values attainable from a c-Si PV cell respectively. The data samples are discussed in detail in the following subsections, providing a summary and evaluation of the optical and electrical performance of these PV cells in relation to various photon management approaches.

*3.1. Front Side Structures*

Various methods have been utilized to improve the transparency of the photovoltaic cells on the front side, such as applying an antireflective coating (ARC), texturing the surface, and utilizing resonators. A graphical representation of these techniques is presented in Figure 3(a).

Single layer dielectric antireflective coating (ARC) is a straightforward method to reduce reflection losses. Silicon Nitride (SiN$_X$) is commonly used as the coating material [158, 159] due to its lower refractive index ($\approx$ 2) than crystalline silicon ($\approx$ 4). Single layer ARC has the potential to achieve 100% optical transmission at the corresponding wavelength. Due to its simple fabrication process and low cost, dielectric ARC is widely used in commercial PV cells. However, it should be noted that the optical performance of single layer ARC is dependent on wavelength and polarization, and it has limited angular optical response which can be a drawback in practical applications [160].

To broaden the spectral and angular responses of ARC, various strategies have been examined. These methods can currently be divided into two categories: patterning on layers with different materials and direct patterning on the silicon wafer (surface texturing). In these architectures, nano or microstructures with gradually varying geometries are positioned above the silicon wafer, resulting in higher transmission. Unlike traditional single layer ARC, these approaches exhibit an omnidirectional and broadband optical response due to the graded index distribution and thus destructive interference of a wide range of wavelengths, which is crucial for PV cells.

Antireflective nanostructures with different shapes have been studied, including nanowires, nanopillars, nanodomes, nanospheres, nanorods, and nanocones. For example, Jeong *et al.* investigated the performance of an ultra-thin silicon nanocone PV cell [161]. In this study, Si nanocones were fabricated on top of a Si substrate, and interdigitated Al contacts were placed on the rear side to prevent shadowing losses. The cell demonstrated 80% external quantum efficiency (EQE) in the 400-800 nm wavelength range, and a $J_{SC}$ of 29 $mA/cm^2$. Here, in addition to the antireflection effect, enhanced light scattering from these structures also contributed to the light trapping. Cordaro *et al.* nanopatterned the SiN$_X$ ARC layer, which provided with a boost of 2.3 $mA/cm^2$ in the $J_{SC}$ in comparison with the planar ARC [162]. Nanowires are another effective geometry in light management. Lee *et al.* reported a radial junction PV cell with hybrid silicon wire structures [163]. In this configuration, p-n junctions were formed along the radius of silicon microwire arrays. The n-type microwires were deposited on a thick silicon wafer while p-type nanowires were deposited on top of them. This design shortened the traveling path of the charge carriers, thus reducing recombination losses. The combined effects of graded index structures and light trapping between the wires significantly reduced reflection losses, resulting in an excellent 97% absorption of light and 39.5 $mA/cm^2$ $J_{SC}$.

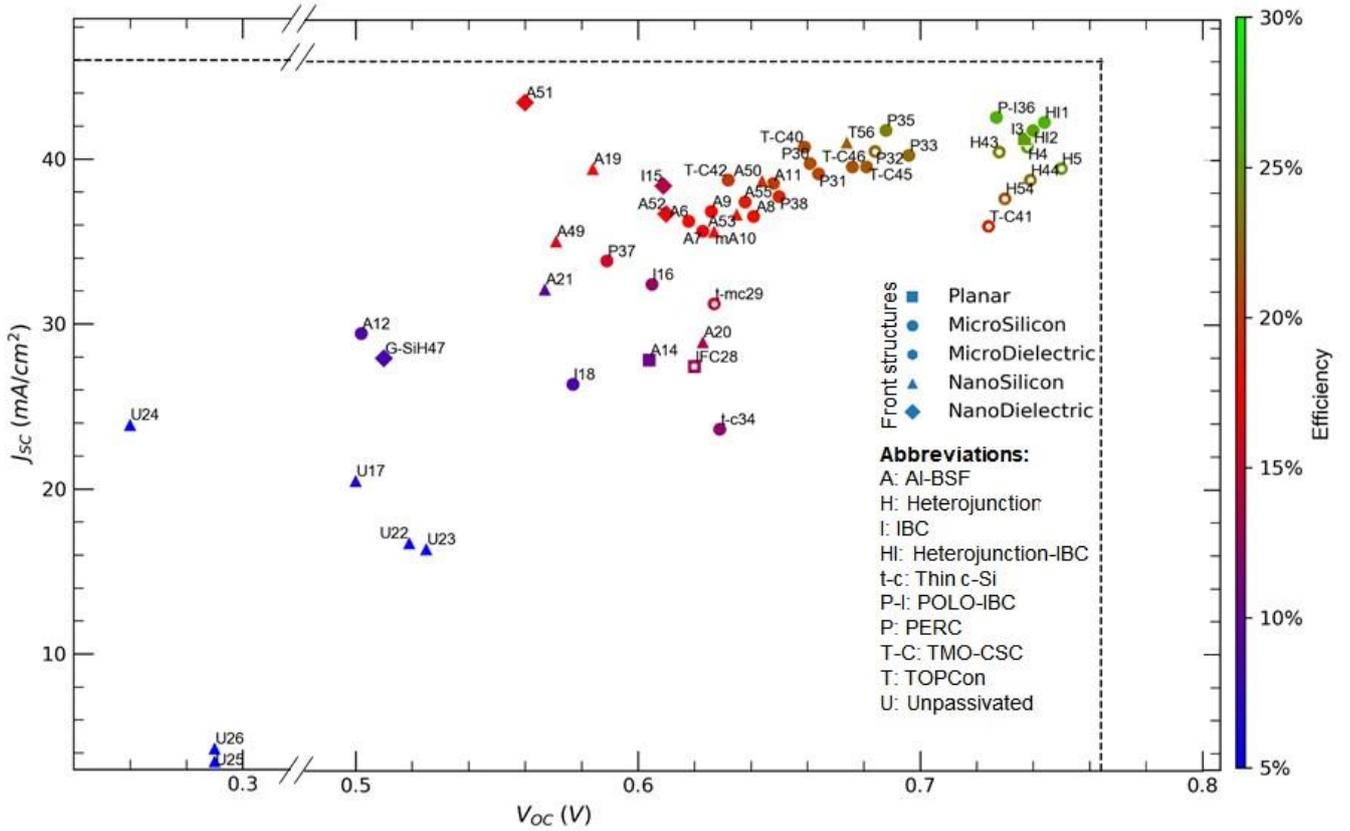

Figure 2: Impact of front side photon management structures and cell types on the short-circuit current density ($J_{SC}$), open-circuit voltage ($V_{OC}$), and efficiency of silicon photovoltaic cells. The horizontal and vertical dotted lines represent the highest achievable $J_{SC}$ and $V_{OC}$ values, respectively. Performance parameters for each cell are provided in Table 1, and the data point annotations in the figure correspond to the cell types and the serial numbers in that table.

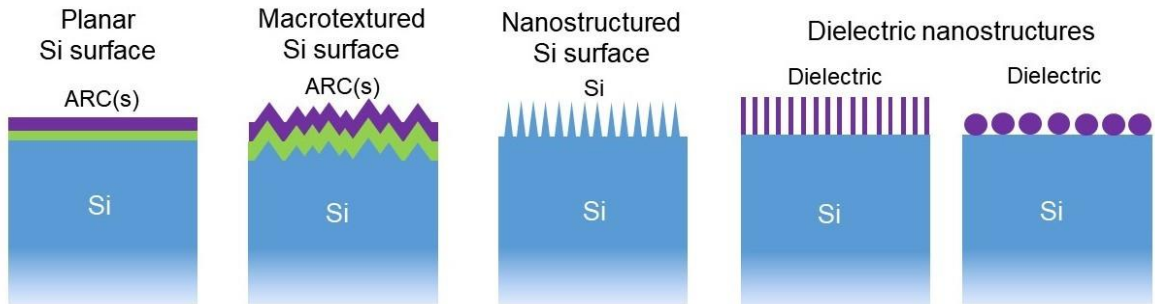

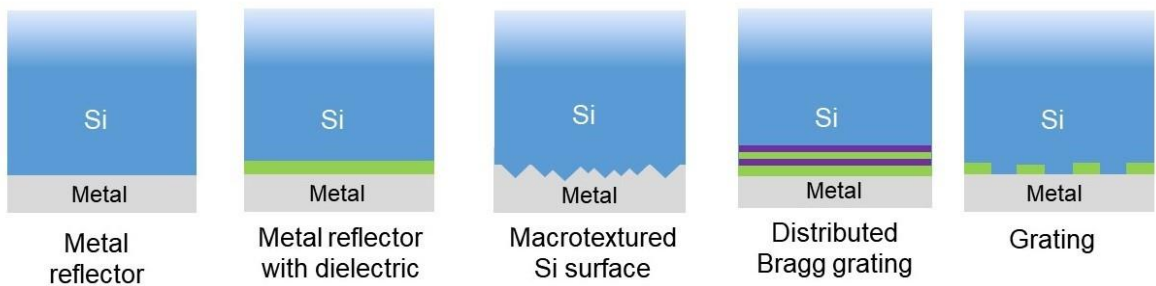

Figure 3: Schematic representation of different photon management structures employed on the front and rear sides of a silicon solar cell (reproduced from [157]).

Surface texturing is a widely used method of reducing front-side reflection losses. The main difference between surface texturing and antireflective nanostructure is that surface texturing is directly etched or deposited on the substrate, without adding a new layer with different materials. Therefore, surface texturing has a simple fabrication process. Similar to antireflective nanostructures, enhanced transmission into the cell is achieved through altered internal angles and graded refractive index caused by gradually varying geometry. For example, Kafle *et al.* [164] reported a nanotextured multicrystalline (mc-Si) PV cell achieving 36.7 $mA/cm^2$ $J_{SC}$. Fellmath *et al.* reported another Al-BSF cell with pyramid texturing on the front surface [165] which achieved an improved $J_{SC}$ of 38.6 $mA/cm^2$. A cost-effective form of nanostructured silicon called black silicon (b-Si) due to its extremely low reflection, is another candidate for excellent light trapping. It can maintain a very high transparency characteristics along a wide wavelength range, especially in the visible and infrared region, resulting in a high efficiency. Savin *et al.* reported a b-Si PV cell with no front-side contact but interdigitated back-contact (IBC) that mitigated reflection losses on the front side [166]. Due to the needle-like silicon textures with high aspect ratio, an effective medium was formed resulting in a huge reduction in the front-side reflection loss. The cell achieved a very high $J_{SC}$ of 42.2 $mA/cm^2$.
Other applications of b-Si include ultra-thin IBC cells [167], and TOPCon cells with passivated b-Si [168].

An interesting approach of improving photon absorption in the device involves the use of optical resonators. When nanoparticles are embedded inside the PV cell, the field distribution becomes highly confined in the active layer which enhances the absorption of the incident photons. However, unwanted plasmonic absorptions can be induced by metallic nanoparticles [169]. Various materials have been adopted as nano resonators, including dielectric (Mie resonance) and metallic (plasmon resonance) materials. By applying particles with different sizes, multiple resonant peaks are excited, resulting in a broadband optical response. Additionally, these nanoparticles also scatter light, extending the optical path length inside the active region. For instance, Yao *et al.* [170] reported a nanoshell
PV cell where a layer of silicon nanospheres helped the transfer of light to the cell through whispering-gallery modes (WGMs) [171, 172]. The cell demonstrated a $J_{SC}$ of 20.1 $mA/cm^2$.

*3.2. Rear Side Structures*

In the absence of a rear side photon management strategy, the light that was not fully absorbed by silicon may exit the cell, leading to a lower $J_{SC}$. A simple way of redirecting these photons to the active region involves using a planar metal contact (similar to the illustration in Figure 3(b): Metal reflector) that reflects light in addition to collecting the photogenerated carriers to the external circuit. However, when incident light falls perpendicularly on a planar metal contact (e.g., Al), it is reflected perpendicularly, resulting in a path length enhancement factor of only two. Instead, if the incident light falls on the planar rear side at an angle (as in Figure 1), which is the case

Table 1: Cell Performance Parameters

| SN | Front Structure | Rear Structure | Thickness of Si ($\mu$m) | Cell Type | $V_{OC}$ (V) | $J_{SC}$ (mA/cm$^2$) | FF (%) | Efficiency (%) | Reference |
|---|---|---|---|---|---|---|---|---|---|
| 1 | MicroSi | Planar | 165 | Heterojunction-IBC | 0.744 | 42.3 | 83.57 | 26.3 | [173] |
| 2 | MicroSi | Planar | 150 | Heterojunction-IBC | 0.74 | 41.8 | 82.76 | 25.6 | [174] |
| 3 | MicroSi | Planar | 130 | IBC | 0.737 | 41.3 | 82.79 | 25.2 | [175] |
| 4 | MicroSi | MicroSi | 160 | Heterojunction | 0.738 | 40.8 | 83.36 | 25.1 | [176] |
| 5 | MicroSi | MicroSi | 98 | Heterojunction | 0.75 | 39.5 | 83.37 | 24.7 | [177] |
| 6 | MicroSi | Planar | 180 | Al-BSF | 0.618 | 36.3 | 79.34 | 17.8 | [159] |
| 7 | MicroSi | Planar | 150 | Al-BSF | 0.623 | 35.7 | 75.53 | 16.8 | [158] |
| 8 | MicroSi | Planar | 150 | Al-BSF | 0.641 | 36.6 | 77.58 | 18.2 | [158] |
| 9 | MicroSi | Planar | 150 | Al-BSF | 0.626 | 36.9 | 75.76 | 17.5 | [158] |
| 10 | NanoSi | Planar | 195 | mAl-BSF | 0.627 | 35.7 | 80.41 | 18 | [164] |
| 11 | MicroSi | Planar | 225 | Al-BSF | 0.648 | 38.6 | 80.36 | 20.1 | [165] |
| 12 | MicroSi | Planar | 75 | Al-BSF | 0.502 | 29.5 | 62.39 | 9.24 | [178] |
| 13 | Planar | NanoDi | 0.450 | Thin-aSi | 0.83 | 14.07 | 56.08 | 6.55 | [179] |
| 14 | Planar | Planar | 280 | Al-BSF | 0.604 | 27.9 | 62.31 | 10.5 | [180] |
| 15 | NanoDi | Planar | - | IBC | 0.609 | 38.45 | 59.96 | 14.04 | [181] |
| 16 | MicroSi | Planar | - | IBC | 0.605 | 32.48 | 65.19 | 12.81 | [181] |
| 17 | NanoSi | Planar | 10 | Unpassivated | 0.5 | 20.59 | 69.84 | 7.19 | [182] |
| 18 | MicroSi | Planar | - | IBC | 0.577 | 26.42 | 58.84 | 8.97 | [181] |
| 19 | NanoSi | Planar | 200 | Al-BSF | 0.584 | 39.5 | 76.29 | 17.6 | [163] |
| 20 | NanoSi | Planar | 10 | Al-BSF | 0.623 | 29 | 75.83 | 13.7 | [161] |
| 21 | NanoSi | Planar | - | Al-BSF | 0.567 | 32.2 | 52.09 | 9.51 | [183] |
| 22 | NanoSi | Planar | 20 | Unpassivated | 0.519 | 16.82 | 60.71 | 5.3 | [184] |
| 23 | NanoSi | Planar | 8 | Unpassivated | 0.525 | 16.45 | 55.93 | 4.83 | [184] |
| 24 | Planar | Planar | - | Unpassivated | 0.26 | 23.9 | 54.71 | 3.4 | [185] |
| 25 | Planar | Planar | 0.225 | Unpassivated | 0.29 | 3.5 | 49.26 | 0.5 | [186] |
| 26 | Planar | Planar | - | Unpassivated | 0.29 | 4.28 | 37.06 | 0.46 | [187] |
| 27 | NanoSi | NanoSi | 0.280 | Thin-aSi | 0.75 | 17.5 | 44.95 | 5.9 | [188] |
| 28 | Planar | MicroSi | 675 | IFC | 0.62 | 27.5 | 77.42 | 13.2 | [189] |
| 29 | MicroSi | NanoSi | 20 | Thin-mcSi | 0.627 | 31.3 | 77.45 | 15.2 | [190] |
| 30 | MicroSi | Planar | - | PERC | 0.661 | 39.8 | 80.58 | 21.2 | [191] |
| 31 | MicroSi | Planar | - | PERC | 0.664 | 39.18 | 79.95 | 20.8 | [192] |
| 32 | MicroSi | MicroSi | - | PERC | 0.684 | 40.54 | 81.54 | 22.61 | [193] |
| 33 | MicroSi | Planar | - | PERC | 0.696 | 40.3 | 81.28 | 22.8 | [58] |

| SN | Front Structure | Rear Structure | Thickness of Si ($\mu m$) | Cell Type | $V_{OC}$ (V) | $J_{SC}$ ($mA/cm^2$) | FF (%) | Efficiency (%) | Reference |
|---|---|---|---|---|---|---|---|---|---|
| 34 | MicroSi | Planar | - | TMO-CSC | 0.629 | 23.7 | 82.51 | 12.3 | [194] |
| 35 | MicroSi | Planar | 180 | PERC | 0.6879 | 41.81 | 82.85 | 23.83 | [195] |
| 36 | MicroSi | Planar | 290 | POLO-IBC | 0.727 | 42.6 | 84.27 | 26.1 | [196] |
| 37 | MicroSi | Planar | 10 | PERC | 0.589 | 33.9 | 78.63 | 15.7 | [197] |
| 38 | MicroSi | Planar | 43 | PERC | 0.65 | 37.8 | 77.74 | 19.1 | [198] |
| 39 | MicroSi | Planar | 180 | TMO-CSC | 0.659 | 40.8 | 79.22 | 21.3 | [199] |
| 40 | MicroSi | MicroSi | 200 | TMO-CSC | 0.72422 | 36 | 74.03 | 19.3 | [200] |
| SN | Front Structure | Rear Structure | Thickness of Si ($\mu m$) | Cell Type | $V_{OC}$ (V) | $J_{SC}$ ($mA/cm^2$) | FF (%) | Efficiency (%) | Reference |
| 41 | MicroSi | Planar | 200 | TMO-CSC | 0.632 | 38.8 | 81.97 | 20.1 | [201] |
| 42 | MicroSi | MicroSi | 230 | Heterojunction | 0.728 | 40.5 | 81.09 | 23.91 | [202] |
| 43 | MicroSi | MicroSi | 150 | Heterojunction | 0.739 | 38.8 | 80.60 | 23.11 | [202] |
| 44 | MicroSi | Planar | 155 | TMO-CSC | 0.681 | 39.6 | 80.84 | 21.8 | [203] |
| 45 | MicroSi | Planar | 175 | TMO-CSC | 0.676 | 39.6 | 80.69 | 21.6 | [204] |
| 46 | NanoDi | Planar | 20 | Gr-Si Heterojunction | 0.51 | 28 | 61.62 | 8.8 | [205] |
| 47 | NanoDi | Planar | > 0.032 | Thin-aSi | 0.885 | 9.24 | 60.90 | 4.98 | [206] |
| 48 | NanoDi | Planar | 50 | Al-BSF | 0.571 | 35.11 | 79.21 | 15.88 | [207] |
| 49 | NanoSi | Planar | - | Al-BSF | 0.644 | 38.77 | 79.30 | 19.8 | [208] |
| 50 | NanoDi | Planar | - | Al-BSF | 0.56 | 43.5 | 68.14 | 16.6 | [209] |
| 51 | NanoDi | Planar | - | Al-BSF | 0.61 | 36.75 | 74.90 | 16.79 | [210] |
| 52 | NanoSi | Planar | 200 | Al-BSF | 0.635 | 36.74 | 79.81 | 18.62 | [211] |
| 53 | MicroSi | MicroSi | 200 | Heterojunction | 0.73 | 37.67 | 79.42 | 21.84 | [212] |
| 54 | MicroSi | Planar | 200 | Al-BSF | 0.638 | 37.47 | 78.93 | 18.87 | [213] |
| 55 | NanoSi | Planar | - | TOPCon | 0.674 | 41.1 | 80.50 | 22.3 | [214] |
| 56 | NanoSi | Planar | 170 | TOPCon | 0.696 | 41.44 | 81.69 | 23.55 | [168] |
| 57 | MicroSi | MicroSi | - | TOPCon | 0.717 | 40.57 | 84.52 | 24.58 | [215] |
| 58 | MicroSi | MicroSi | 130 | Heterojunction | 0.751 | 41.45 | 86.07 | 26.8 | [90] |
| 59 | MicroSi | Planar | - | PERL | 0.706 | 42.7 | 82.8 | 25.0 | [216] |

Elaboration: MicroSi=microscale silicon, NanoSi=nanoscale silicon, NanoDi=nanoscale dielectric.

when a textured front side photon management technique is used, the rear side also reflects at an angle. With that, the unabsorbed photons keep bouncing back and forth between the front and the rear side of the cell, ensuring a big boost in the path length enhancement factor. However, the unpassivated metal-semiconductor boundary acts as a recombination center which annihilates some of the photogenerated carriers before being collected to the external circuit, lowering $J_{SC}$ and $V_{OC}$. Additionally, the metal contact in the rear side can also cause parasitic optical absorption loss, especially in the longer wavelengths, hurting the performance further [64, 145]. One solution of this problem is to form an Al-Si eutectic, (also called back surface field or BSF) which keeps the minority carriers away from the high-recombination rear surface, also acts as a reflector. Several versions of the Al-BSF cells can be found in the literature: full-area Al-BSF, local Al-BSF, and special structures. For instance, Vermang *et al.* [158] reported full-area Al-BSF, blistered Al-BSF and local Al-BSF demonstrating 35.7, 36.9 and 38.7 $mA/cm^2$ $J_{SC}$ respectively. The improvement in performance in these cells came from the enhanced rear-side reflection and improved surface passivation. In PERC cells, although the the main purpose of the dielectric layer (i.e., $Al_2O_3$) in the rear-side metalsemiconductor boundary was surface passivation, it also helps increase the rear-side reflection [217] by reducing the unwanted metallic-absorption loss in the longer wavelengths [64, 145]. For example, Hannebauer *et al.* reported a commercial PERC cell structure [218] with a $J_{SC}$ of 39.8 $mA/cm^2$. The rear side photon management technique of all the cell falls into the category of planar metal reflector with dielectric in Figure 3(b).

When it is very common to have a planar rear side (like most of the cells in table 1), other type of structures are also used for the photon management in the rear side. Microscale structures, nanoscale structures (e.g., nanocones, nanodomes), distributed Bragg grating, and 2-D grating are examples of such structures. For instance, Zhu *et al.*, used nanocone back reflectors in a-Si:H PV cells [188]. In this study, a-Si nanocones in the scale of hundreds of nanometers were deposited on the substrate with the same material, forming periodic nanodome structures. This structure provided scattering which helped enhance the optical path length and improved absorption in the active region. The design yielded a $J_{SC}$ of 17.5 $mA/cm^2$. Also, in a recent work, Cordaro *et al.* used nanopatterned metal reflector engineered to suppress plasmonic loss and promote scattering [219].

Grating structures have been used on the rear side of a cell due to their ability to support resonant modes and scatter light. For example, Zeng *et al.* [189] reported a hybrid grating-photonic crystal back reflector silicon PV cell. The design included a reflection grating on the rear side of the cell and a 1D distributed back reflector (DBR) layer beneath the grating. The DBR achieved over 99.8% reflectance in the wavelength range of 800-1100 nm. This design resulted in a $J_{SC}$ of 27.5 $mA/cm^2$.

Resonant modes have also been shown to improve light trapping on the rear side of a PV cell. For instance, Tu *et al.* reported the use of a double wall carbon nanotubes (DWCNTs) in amorphous silicon (a-Si) PV cells [179]. The DWCNTs were spin-coated on Ti/Ag back contacts to excite plasmon resonances and enhance light scattering in the range of 589-700 nm. The reported $V_{OC}$ and $J_{SC}$ were 0.83 V, and 14.07 $mA/cm^2$ respectively, yielding an efficiency of 6.55%

*3.3. Photon Management Structures Not Yet Incorporated in a Practical PV cell*

Recent literature presents various photon management structures grown on isolated substrates or simulated, such as Heidarzadeh *et al.*'s idea for increasing photogenerated current in thin-film PV cells [220] using hemispherical coreshell nanoparticles on the front and triangular grating reflectors on the rear which achieved $J_{SC}$ of 22 $mA/cm^2$. The forward scattering of the core-shell nanoparticles is a positive aspect for $J_{SC}$ but they incur plasmonic absorption, causing dips in the active material's absorption spectra and not fully utilizing the incident sunlight. Comparing this to our SunSolve simulation of planar Si and no ARC (1-R characteristics in Figure 4(a)), it showed better absorption in Si for some wavelengths but poorer in others. However, it under-performs compared to our planar Si and ARC simulation. Therefore, using these core-shell nanoparticles does not significantly improve cell performance.

In a similar design, trapezoidal pyramid nanostructure (i.e., nanosilicon) are proposed for the front side and inverted pyramid nanotexture (i.e., nanosilicon) for the rear side of a 0.9 $\mu$m thick c-Si/ZnO heterojunction PV cell [221]. The simulations resulted in a $J_{SC}$ of 41.94 $mA/cm^2$, with a front reflectance of ≤5% for 400 - 1000 nm wavelength and higher for other wavelengths. Although the results are promising, it is challenging to implement the nano-trapezoidal Si texture on the front side and passivate both sides, particularly the rear side where Si inverted pyramids are filled with Al.

Recently, Wang *et al.* has developed a ZnO nano-needle array (i.e., nanodielectric structure) for front-side photon management [222] on a textured Si substrate, which has been experimentally shown to reduce front reflectance to ≤8% at 400-1000 nm wavelength. This structure achieved a $J_{SC}$ of 37.8 $mA/cm^2$, an increase of 1.8 $mA/cm^2$ compared to textured Si front surface. The advantage of this structure is that it avoids the challenge of passivating a nanostructured Si while providing similar front reflectance. Another similar front structure is AZO (Al doped ZnO) nanorods which experimentally provided a front surface reflectance of about 12% at 400-900 nm wavelength [223].

Saravanan *et al.* designed a 1-D photonic crystal (i.e., a distributed Bragg grating) for the rear side of an a-Si PV cell [224], which consisted of alternating layers of Si and $SiO_2$ and had the potential to make the rear side reflectance near perfect. However, it is unclear how rear side contacts could be formed through this dielectric reflector.

Hossain *et al.* experimentally demonstrated the potential of self-assembled $Al_2O_3$ nanostructures for both surface passivation and photon management in the rear side, achieved through suppression of parasitic optical loss in the metal contacts [64, 157, 225]. With rear side reflectance as high as 95%, these structures could bring up to 2.9 $mA/cm^2$ increase in $J_{SC}$. In a similar work by Shameli, a polarization independent phase gradient metasurface showed [226] 83% rear side reflectance in simulation. However, surface recombination at the Si/Al boundary was not addressed, which could potentially negatively impact the cell performance.

Dhawan *et al.* proposed using transformation optics to design photon management structures on Si surfaces, recognizing that texturing the surface increases recombination and it may not be possible to fully passivate it [227]. They proposed using optically equivalent planar layers with varying refractive indices instead of a nano-structured surface. However, it remains uncertain how to fabricate these precise planar layers with varying refractive indices.

Sun *et al.* suggested using metallic light-trapping electrodes on the front side of the PV cells, which can significantly decrease shadowing loss and have potential for use in the front-side metal grids, increasing the active area of the cell [228–230].

Khokhar *et al.* recently designed a TOPcon cell that incorporated a nanosilicon structure on both sides of the cell [231]. By considering both bulk and surface recombination in their simulation, they were able to achieve a $J_{SC}$ of 43.5 $mA/cm^2$, a $V_{OC}$ of 0.762 V and an *FF* of 83%. These results are slightly higher than those of practical TOPcon cells reported in Table 1, suggesting that there is potential for further improvement in the performance of practical TOPcon cells through improved photon management and advancements in processing techniques that can reduce recombination losses.

*3.4. Combined Analysis of the Front and Read Side Photon Management Structures*

In a PV cell, both front and rear side photon management structures are needed in order to harness the full potential of the incident light, and together, they contribute to an increase in the $J_{SC}$ of the cell. However, other factors in the cell architecture like cell thickness and surface passivation also contribute to the $J_{SC}$ value. Particularly, surface passivation can impact $J_{SC}$, $V_{OC}$, and *FF*. Therefore, when we compare the contribution of the photon management structures in terms of $J_{SC}$, the contribution from the surface passivation needs to be decoupled. It is helpful to discuss the photon management structures of the same cell type altogether in this regard. Figure 2 presents the performance parameters of various recently made Si PV cells, as well as their corresponding photon management techniques and cell types; the cells have been chosen in such a way that they represent a wide variety of photon management techniques and cell types, helping us evaluate their impacts on cell performance. The maximum attainable $V_{OC}$ and $J_{SC}$ from a c-Si PV cells are shown as dotted lines which help determine how far a particular cell is standing from its optimum performance and point to the possible reasons behind that. Table 1 provides additional information on these cells i.e., cell thickness and *FF*.

The importance of surface passivation is evident in the unpassivated PV cells represented by the data points U17, U22, and U23, which have nanosilicon structures on the front side and planar rear side. These arrangement provided the $J_{SC}$ between 12-21 $mA/cm^2$. However, as it was challenging to passivate nanosilicon structures, the poor passivation of the front nanostructures resulted in a high recombination loss and very low efficiency. Additionally, U25 and U26 are Si nanowire-based PV cells with no notable photon management structures and surface passivation which caused low $V_{OC}$, $J_{SC}$ and *FF*, and extremely low efficiency.

Many data points represent Al-BSF cells where the BSF layer acts as both a passivation layer and reflector. A variety of structures are employed on the front side, including planar, microsilicon, nanosilicon, and nanodielectric structures for photon management. The highest efficiency Al-BSF cell (A11, 20.1% efficiency) has a microsilicon front structure, providing a $J_{SC}$ of 38.6 $mA/cm^2$ with a Si thickness of 225 $\mu m$. The next highest efficiency Al-BSF cell (A49, 19.8% efficiency) has a $J_{SC}$ of 38.77 $mA/cm^2$. It is important to note that front-side nanostructures have slightly higher $J_{SC}$ values but slightly lower $V_{OC}$ and *FF* values in the case of A49, likely due to poor surface passivation of the nanostructures. The lowest efficiency Al-BSF cell (A12, 9.24% efficiency) has a microsilicon front structure, poor performance in terms of $V_{OC}$, $J_{SC}$, and *FF*, indicating high recombination. Additionally, in this cell, the 75 $\mu m$ thickness was not sufficient for capturing incident photons, contributing to a lower $J_{SC}$.

PERC cells are currently the dominant PV cells in the global market [63, 64]. These cells have passivated front and rear sides, reducing surface recombination and achieving high $V_{OC}$, $J_{SC}$ and *FF* values with the same photon management as Al-BSF cells. For instance, P47 achieved 21.2% efficiency with a $V_{OC}$ of 0.627 V, $J_{SC}$ of 39.8 $mA/cm^2$ and *FF* of 80.58% using a microsilicon front and planar rear. Structuring the rear side while maintaining good surface passivation improves performance further, as seen in cell P32, which has microsilicon structures on both sides and achieved a $V_{OC}$ of 0.684 V, $J_{SC}$ of 39.8 $mA/cm^2$, *FF* of 81.54%, and efficiency of 22.61%

The TOPcon cell [83] is a new type of PV cell that significantly reduces surface recombination by utilizing a thin tunneling oxide to passivate the contact, keeping the contact resistivity of the cell almost unchanged. In one of these cells (T55), nanosilicon front side and planar rear provided 41.1 $mA/cm^2$ $J_{SC}$, pointing to a good photon management arrangement. This cell achieved a $V_{OC}$ of 0.674 V and an efficiency of 22.3%. However, it was found that the surface passivation of the nanosilicon structure in T55 was not sufficient, prompting the introduction of a new nanosilicon structure in T56 that reduces the difficulty of passivation. This new structure, which involves forming a random pyramid texture and then creating nanoscale pores through reactive ion etching, resulted in increased photocurrent density, $V_{OC}$ and efficiency compared to T55. Another TOPcon cell, T57, used a microsilicon structure on both sides, but did not achieve as high photocurrent density as T55 and T56 due to its relatively higher front reflectance. However, the passivation quality of the microsilicon structure was better, resulting in a higher $V_{OC}$ and efficiency. These examples demonstrate that increasing the photogenerated current does not always improve cell performance and a comprehensive analysis should be conducted when implementing a photon management approach.

The heterojunction cells in the dataset have achieved relatively higher efficiency by using microsilicon structures on both the front and rear sides. This high performance was achieved due to effective photon management, selecting the appropriate cell thickness, and proper surface passivation. For example, H4 demonstrated a $V_{OC}$ of 0.738 V, a photocurrent density of 40.8 mA/cm2, and a *FF* of 83.36%, resulting in an efficiency of 25.1%. Its performance was surpassed by H2, which achieved an efficiency of 25.6% by using a heterojunction-IBC structure. H2 had a $V_{OC}$ and *FF* that were similar to H4, but the main improvement came from an increase in photocurrent density (41.8 $mA/cm^2$) due to the adoption of an IBC structure on the rear side which also trapped some light to the cell.

While it is common to have a textured front and a planar rear side in solar cells, there are also cases where the opposite is used. For instance, the cell IFC28 had a planar front side and a photonic crystal back reflector with a grating structure on the rear side. This arrangement resulted in a photocurrent density of 27.5 $mA/cm^2$. The $V_{OC}$ and *FF* of this cell were 0.62 V and 77.42%, respectively. However, the photocurrent density could have been further improved if the front side also had a photon management scheme. Additionally, the cell thickness was 675 $\mu m$, a value much greater than the diffusion length [232, 233] of silicon (100-300 $\mu m$) [97, 234, 235]. This resulted in some nonradiative recombinations, the absence of which could have led to even higher $V_{OC}$, $J_{SC}$, and *FF*.

A number of cells were investigated with nanodielectric structures on the front side, which provided some level of light trapping compared to cells with no structure on the front. However, these schemes failed to achieve significant efficiency. For example, GSiH46 achieved a $V_{OC}$ of 0.51 V, a photocurrent density of 28 $mA/cm^2$, and a *FF* of 61.62% using a bilayer graphene on the front side and a planar rear side. The front reflectance of the cell was less than 5% for the wavelength range of 520-700 nm and less than 10% for longer wavelengths, indicating that the cell was able to capture a good number of photons, yet failed to achieve high efficiency. There are several reasons for this. Firstly, it used whispering gallery mode [171, 172] to confine incident light to the cell. However, in this arrangement, perpendicularly incident light does not see any redirection by the front side, as is usually seen for pyramid textured front surfaces (microsilicon structure) [93, 94, 236]. Secondly, the cell thickness of only 20 $\mu m$ was not sufficient for adequate light trapping. Additionally, the lower values of $V_{OC}$ and *FF* also indicate that the surface passivation quality was poor.

## 4. Practical Limit of Photon Management and Improvement Pathways

Quantifying the contribution of each technique is crucial in the pursuit of the most effective photon management approach. According to the discussions in the previous section, microsilicon and nanosilicon structures showed the greatest potential for use as front-side photon management structures. Furthermore, when rear side structures such as microsilicon are used, they reflect incident light at an angle which enhances path length further. To determine the highest achievable photocurrent density practically in these configurations, we conducted a study using SunSolve [237]. Pyramid structures were used as a representative of microsilicon structures and black silicon was used as a representative of nanosilicon structures for the front side. On the rear side, both metallic and dielectric rear sides with and without full scattering properties were investigated. Additionally, the impact of cell thickness in the presence of front and rear side photon management mechanisms was also studied.

Figure 4(a) illustrates the performance of different front side photon management structures with a fixed rear side consisting of 1 $\mu$m Al-Si + 10 nm $Al_2O_3$ + 100 nm $SiN_X$, with no scattering (similar to the rear side structure in Figure 3(b): metal reflector with dielectric). The cell thickness chosen for this study was 175 $\mu$m, a value consistent with the thickness observed in an industrial Si PV cell. The study considers three different front structures: a planar surface, a random pyramid texture, and black Si (5 $\mu$m tall, 87.5° sidewall angle), both with and without

75 nm $SiN_X$ on them; $SiN_X$ worked as an ARC. The first part of Figure 4(a) shows the 1-R (total absorption) and EQE characteristics, where R is the total reflectance of the cell.

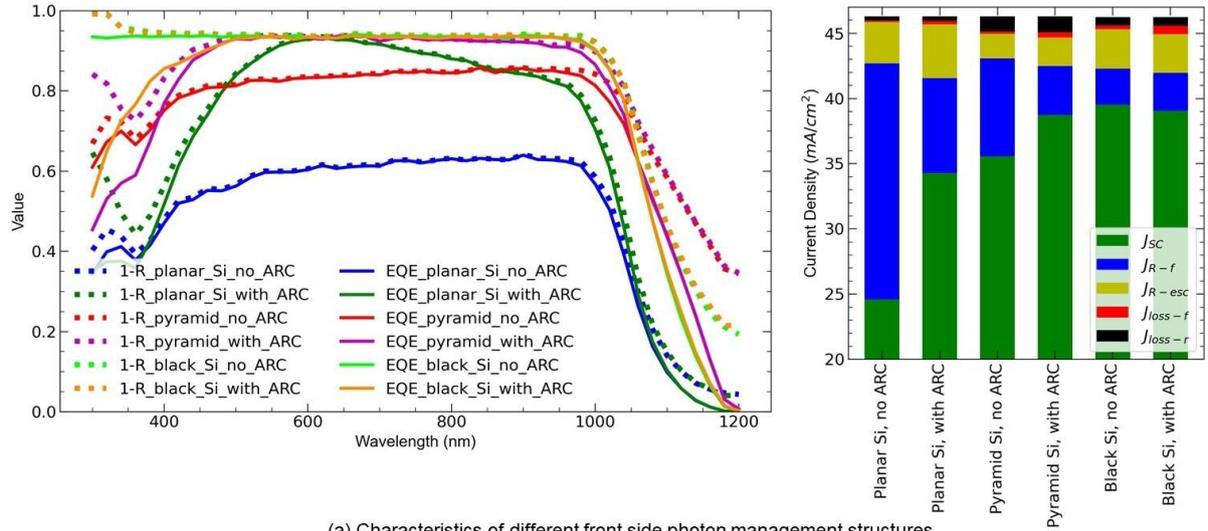

(a) Characteristics of different front side photon management structures
(fixed rear side: PERC - 1um AlSi+10nm $Al_2O_3$+100 nm $SiN_X$, no scattering. Wafer thickness: 175 μm)

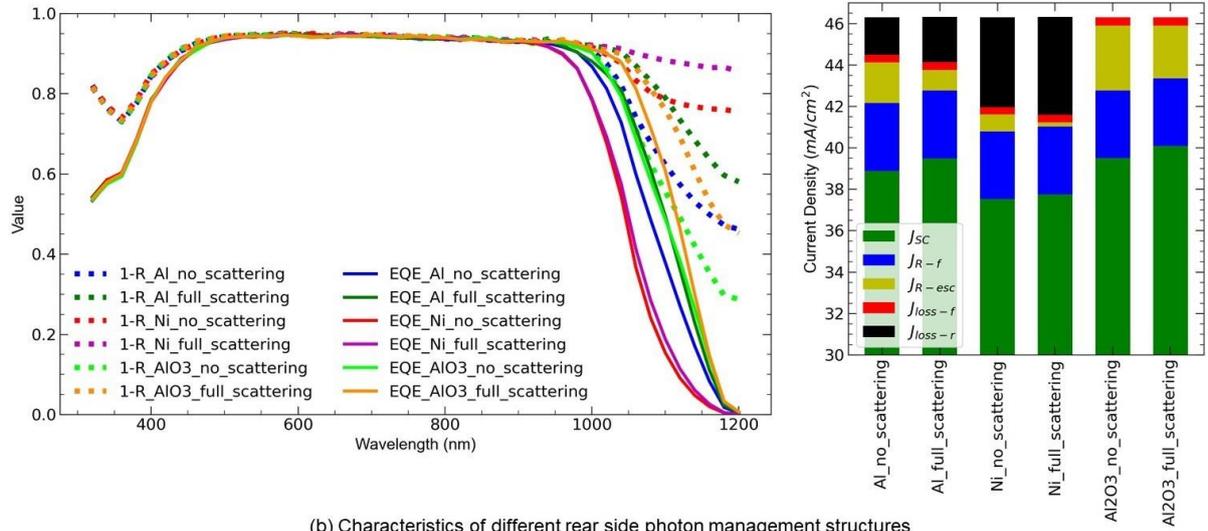

(b) Characteristics of different rear side photon management structures
(fixed front side: 75nm $SiN_X$ + pyramid texture. Wafer thickness: 175 μm)

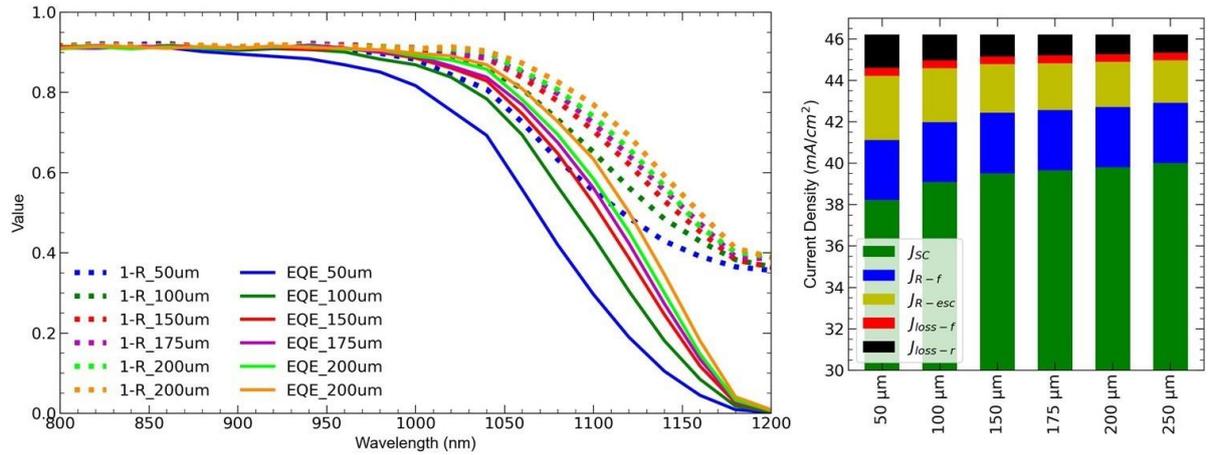

(c) Impact of cell thickness on photon management (Front: 5 μm high random pyramid, 87.5° (black Si), no ARC
Rear: PERC - 1 μm AlSi contact + 10 nm $Al_2O_3$ + 100 nm $SiN_X$, full scattering)

Figure 4: Assessment of the impacts of photon management structures on cell performance. Legend: $J_{SC}$: short-circuit current density, $J_{R-f}$: loss in current density due to front reflectance, $J_{R-esc}$: loss in current density due to escape reflectance, $J_{loss-f}$: loss in current density (absorption) at the front side, and $J_{loss-r}$: loss in current density (absorption) at the rear side.

It is important to note that 1-R accounts for both the absorption in the active region (desired) and the parasitic absorption loss in the front and rear side of the cell (undesired) [57, 95]. We observed that a PV cell with a planar silicon surface and no ARC on the front side was able to absorb approximately 60% of the incident light in the 600-1000 nm wavelength range, with a decrease in absorption for other wavelengths. A slight difference between the EQE and 1-R was noticed in the shorter (<400 nm) and longer (>1100 nm) wavelength ranges, which can be attributed to the parasitic optical absorption on the front and rear sides of the cell respectively. The addition of a 75 nm $SiN_X$ ARC on the planar front side significantly improved the 1-R and EQE of the cell. Additionally, incorporating pyramid textures on the front side also greatly improved the 1-R and EQE compared to a planar surface without an ARC, but exhibited a lower performance in the 430-820 nm range. The highest EQE and 1-R values were achieved using a black silicon front, which was able to absorb approximately 95% of the incident light. Notably, applying an ARC on the black silicon surface resulted in a slight decrease in EQE in the shorter wavelength range, due to parasitic absorption within the ARC. The bar chart in Figure 4(a) breaks down the total incident light energy into various components [57] including photogenerated current density ($J_{SC}$), loss in current density due to front reflectance ($J_{R-f}$) and escape reflectance ($J_{R-esc}$), and loss in current density due to parasitic optical loss on both the front ($J_{loss-f}$) and rear sides ($J_{loss-r}$) of the cell. It indicates that the use of an ARC on planar silicon increased current density by 9 $mA/cm^2$, primarily as a result of reduced front reflectance loss, however it also slightly increased the escape reflectance loss. This is due to the Si-$SiN_X$-air boundary having a lower refractive index contrast than the Si-air boundary, making it easier for light to escape. Using a pyramid silicon surface with an ARC structure further reduced both front and escape reflectance loss. However, using a black silicon structure decreased front reflectance loss and parasitic optical loss on both the front and rear sides of the cell, but increased escape reflectance loss. To further reduce escape reflectance loss, the thickness of the cell could be increased until it reaches the minority carrier diffusion length of silicon, allowing for absorption of longer wavelength photons.

We examined the characteristics of various rear-side photon management structures, all of which had the same front structure consisting of a 75 nm $SiN_X$ ARC on a pyramid texture, with a wafer thickness of 175 $\mu$m. The structures evaluated included planar Al, Al structures designed to scatter light, planar nickel (Ni, a lossy metal with good electrical transport properties), Ni structures designed to scatter light, planar $Al_2O_3$ (which has poor electrical transport properties but excellent surface passivation), and $Al_2O_3$ structures designed to scatter light. Figure 4(b) indicates that all these rear structures displayed similar EQE and reflectance characteristics in the shorter wavelength range, which can be attributed to their consistent front side structure. However, the rear structure primarily influenced the absorption of the longer wavelength photons. The scattering structures generally exhibited higher absorption than their planar counterparts, as more of the unabsorbed light is reflected back and forth between the front and rear sides, increasing the total path length for absorption in silicon. Additionally, the highest absorption in the longer wavelength range was observed with the Ni structure designed to scatter light, followed by the planar Ni. However, these structures also exhibited the lowest EQE, indicating that most of the absorption occurred in Ni (undesired), rather than silicon. The highest EQE in the longer wavelength range was observed in the case of the $Al_2O_3$ structure designed to scatter light, due to the absence of parasitic optical loss on the rear side, which is typically present in metals [64, 145, 146, 238]. The bar chart presented in Figure 4(b) supports the findings that $Al_2O_3$ structures did not exhibit any $J_{loss-r}$, while Ni structures exhibited the highest $J_{loss-r}$. Two key conclusions can be drawn from this information. Firstly, the $J_{SC}$ achieved using the scattering $Al_2O_3$ structure was 40 $mA/cm^2$. In order to increase the $J_{SC}$ further with the same front structure, the cell thickness can be increased, doing so however could lead to increased recombination losses in the bulk. Secondly, $Al_2O_3$ is a dielectric material, using which on the entire rear surface is not allowed as it hinders current collection. A viable solution in this context could involve employing passivating contacts, similar to those used in the TOPCon, HIT or Poly-Si contact architecture.

The impact of the cell thickness on the cell optical performance is illustrated in Figure 4(c). The cell front side is assumed to be black Si without an ARC and the rear side is assumed to be composed of 1 $\mu$m Al-Si, 10 nm $Al_2O_3$, and 100 nm $SiN_X$ with full scattering. As shorter wavelength photons are absorbed in the front side and are not affected by the cell thickness, we presented the 1-R and EQE characteristics only in the 800 - 1200 nm wavelength range. As expected, an increase in thickness was observed to increase the 1-R and EQE, but the rate of increase slowed down with increasing thickness.

In light of the findings discussed above, the dataset presented in this paper (Table 1) was analyzed in more detail. The highest efficiency cell had an efficiency of 26.8%, which was achieved using a microsilicon front, and microsilicon rear side. To enhance its light trapping further, a nanostructured front and micro-structured rear with scattering capabilities could potentially be used. However, passivating a nanostructured surface still remains as a significant challenge. Fortunately, there have been some recent advancements in passivating black Si [239–243] which could potentially be utilized in order to improve the performance of the cell further.

## 5. Conclusion

To minimize the gap between practical efficiency and the Shockley–Queisser limit, it is imperative to incorporate better photon management structures. These structures should exhibit minimal reflectance on the front and high reflectance with scattering properties on the rear side. It is crucial to prevent any contribution to surface recombination or resistive losses. This review comprehensively analyzes various photon management structures, considering $V_{OC}$, $J_{SC}$, $FF$, and efficiency. Many cells underperformed due to the lack of an optimal combination of front and rear structures and inadequate surface passivation. To enhance photon management, it is essential to select an appropriate cell thickness, employ front-side photonic nanostructures reflecting no light across the solar spectrum, implement rear-side structures with near-perfect reflectance and scattering characteristics in longer wavelengths, and ensure effective passivation of both surfaces. Achieving these goals requires careful consideration of cell architecture, material choices, and process technologies.

**Acknowledgments**

This material is based upon work supported by the U.S. Department of Energy's Office of Energy Efficiency and Renewable Energy (EERE) under the Solar Energy Technologies Office Agreement Number DE-EE0007533. The authors would like to acknowledge Prof. Pieter G. Kik for his insightful discussions and valuable perspectives that significantly enhanced the quality of this work. During the preparation of this work, the authors used ChatGPT in order to correct grammatical errors in some parts of the manuscript. After using this tool/service, the authors reviewed and edited the content as needed and take full responsibility for the content of the publication.

**Data Availability**

Datasets related to this article can be found at https://doi.org/10.6084/m9.figshare.23573409.v1, an open source online data repository hosted at figshare [244].